# アプリケーションループ文GPU自動オフロード技術の改善検討

山登庸次†

† NTT ネットワークサービスシステム研究所，東京都武蔵野市緑町 3-9-11
E-mail: †yoji.yamato.wa@hco.ntt.co.jp

**あらまし**　この 4，5 年で，CPU に比べた処理性能，電力効率等の長所から GPU，FPGA 等を利用したシステムが増えている．しかし，GPU，FPGA 等のシステムでの利用には，CUDA や HDL 等のハードウェアを意識した技術仕様の理解が必要であり，ハードルは高い．これらの背景から，私は，プログラマーが CPU 向けに開発したソースコードを，適用される環境に応じて，自動で変換し，リソース量等を設定して，高い性能で運用可能とする環境適応ソフトウェアのコンセプトを提案している．そのコンセプトの一要素として，CPU 向けプログラムの GPU，FPGA への適応について一部アプリケーションに対して自動化を実現し，評価している．本論では，以前の研究の GPU 自動オフロード技術について，より多くのアプリケーションをより高速化するための方式改善を検討し，提案を行う．提案手法を市中のアプリケーションで評価する．

# Improvement of Automatic GPU Offloading Technology for Application Loop Statements

Yoji YAMATO†

† Network Service Systems Laboratories, NTT Corporation, 3-9-11, Midori-cho, Musashino-shi, Tokyo
E-mail: †yoji.yamato.wa@hco.ntt.co.jp

**Abstract**　In recent years, with the slowing down of Moore's law, utilization of hardware other than CPU such as GPU or FPGA is increasing. However, when using heterogeneous hardware other than CPUs, barriers of technical skills such as CUDA and HDL are high. Based on that, I have proposed environment adaptive software that enables automatic conversion, configuration, and high-performance operation of once written code, according to the hardware to be placed. Partly of the offloading to the GPU and FPGA was automated previously. In this paper, I improve and propose a previous automatic GPU offloading method to expand applicapable software and enhance performances more. I evaluate the effectiveness of the proposed method in multiple applications.

## 1. はじめに

近年，CPU の半導体集積度が 1.5 年で 2 倍になるというムーアの法則が減速するのではないかと言われている．そのような状況から，メニーコアの CPU だけでなく，GPU（Graphics Processing Unit）や FPGA（Field Programmable Gate Array）等のハードウェアの活用が増えている．例えば，Amazon 社は，GPU，FPGA 等をクラウドのインスタンス（例えば, [1]- [13]）として提供しており [14]，Microsoft 社は FPGA を使って Bing の検索効率を高めるといった取り組みをしている [15].

しかし，CPU 以外のハードウェアをシステムで適切に活用するためには，ハードウェアを意識した設定やプログラム作成が必要であり，CUDA（Compute Unified Device Architecture）[16]，OpenCL（Open Computing Language）[17]，HDL（Hardware Description Language）といった知識が必要になってくるため，大半のプログラマーにとっては，スキルの壁が高い．

一方，IoT（Internet of Things）技術（例えば, [18]- [21]）は普及してきており，ネットワークにつながるデバイスも，既に数百億と増えており，2030 年には兆台がつながると予測されている．IoT を用いた応用は，医療，流通，製造，農業，エンタメ等に広がっており，サービス合成技術等 [22]- [30] を活用して，

製品が届くまでの過程を可視化するなどの応用がされている．

IoT を用いたシステムで，IoT デバイスを詳細まで制御するためには，組み込みソフトウェア等のスキルが必要になることがある．Raspberry Pi 等の小型端末をゲートウェイ（GW）に，多数のセンサデバイスを集約管理することも頻繁にされるが，小型端末の計算リソースは限定されるため，利用環境に応じて管理の設計が必要となる．

背景を整理すると，CPU 以外の GPU や FPGA 等のハードウェア，多数の IoT デバイスを活用するシステムは今後ますます増えていくと予想されるが，それらを最大限活用するには，壁が高い．そこで，そのような壁を取り払い，CPU 以外のハードウェアや多数の IoT デバイスを十分利用できるようにするため，プログラマーが処理ロジックを記述したソフトウェアを，配置先の環境（GPU，FPGA や IoT GW 等）にあわせて，適応的に変換，設定し，環境に適合した動作をさせるような，プラットフォームが求められている．

Java [31] は 1995 年に登場し，一度記述したコードを，別メーカーの CPU を備える機器でも動作可能にし，環境適応に関するパラダイムシフトをソフト開発現場に起こした．しかし，移行先での性能については，適切であるとは限らなかった．そこで，私は，一度記述したコードを，配置先の環境に存在する GPU や FPGA，IoT GW 等を利用できるように，変換，リソース設定等を自動で行い，アプリケーションを高性能に動作させることを目的とした，環境適応ソフトウェアを提案した．合わせて，環境適応の一要素として，ソフトウェアの GPU，FPGA へのオフロードの自動化をある範囲のアプリケーションに対して実現している．本稿では，以前の研究で進化的計算手法を用いて自動化を行った GPU オフロード技術に対して [32] [33]，より多くのアプリケーションに適用可能とし，より高速化するための方式改善を提案する．提案した手法を実装し，複数のアプリケーションで GPU 自動オフロード技術改善の有効性評価をする．

## 2. 既存技術

環境適応ソフトウェアとしては，Java がある．Java は，仮想実行環境である Java Virtual Machine により，一度記述した Java コードを再度のコンパイル不要で，異なるメーカー，異なる OS の CPU マシンで動作させている（Write Once. Run Anywhere）．しかしながら，移行先で，どの程度性能が出るかはわからず，移行先でのデバッグや性能に関するチューニングの稼働が大きい課題があった（Write Once, Debug Everywhere）．

GPU の並列計算パワーを画像処理でないものにも使う GPGPU（General Purpose GPU）（例えば [34]）を行うための環境として CUDA が普及している．CUDA は GPGPU 向けの NVIDIA 社の環境だが，FPGA，メニーコア CPU，GPU 等のヘテロなハードウェアを同じように扱うための仕様として OpenCL が出ており，その開発環境 [35] [36] も出てきている．CUDA，OpenCL は，C 言語の拡張を行いプログラムを行う形だが，プログラムの難度は高い（FPGA 等のカーネルと CPU のホストとの間のメモリデータのコピーや解放の記述を明示的に行う等）

CUDA や OpenCL に比べて，より簡易にヘテロなハードウェアを利用するため，指示行ベースで，並列処理等を行う箇所を指定して，指示行に従ってコンパイラが，GPU 等に向けて実行ファイルを作成する技術がある．仕様としては，OpenACC [37] や OpenMP 等，コンパイラとして PGI コンパイラ [38] や gcc 等がある．OpenACC は，Fortran/C/C++向けの仕様であるが，Java 向けには，IBM の Java JDK [39] が，Java のラムダ記述に従った GPU オフロード処理を行える．

CUDA，OpenCL，OpenACC 等の技術仕様を用いることで，FPGA や GPU へオフロードすることは可能になっている．しかしハードウェア処理自体は行えるようになっても，高速化することには課題がある．例えば，マルチコア，メニーコア CPU 向けに自動並列化機能を持つコンパイラとして，Intel コンパイラ [40] 等がある．これらは，自動並列化時に，コードの中の for 文，while 文等の中で並列処理可能な部分を抽出して，並列化している．しかし，FPGA や GPU を用いる際は，CPU と FPGA，GPU の間のメモリデータ転送のオーバヘッドのため，並列化しても性能がでないことも多い．FPGA や GPU により高速化する際には，OpenCL や CUDA の技術者がチューニングを繰り返したり，PGI コンパイラ等を用いて適切な並列処理範囲を探索し試行することがされている．

このため，OpenCL や CUDA 等の技術スキルが乏しいプログラマーが，FPGA や GPU を活用してソフトウェアを高速化することは難しいし，自動並列化技術等を使う場合も並列処理箇所の試行錯誤等の稼働が必要だった．

並列処理箇所の試行錯誤を自動化する取り組みとして，著者の以前の研究がある [32] [33]．これら研究は，GPU オフロードに適したループ文を，進化的計算手法を用いて，検証環境での性能測定を繰り返すことで，適切に抽出し，ネストループ文内の変数をできるだけ上位のループで CPU-GPU 転送を一括化することで，自動での高速化を行っていた．しかし，実利用を考えた際に，適用できるアプリケーションが限られることや，CUDA を使った手動高速化に比べて性能改善が物足りないことが課題として挙がっていた．

## 3. ループ文の GPU 自動オフロード手法の改善

### 3.1 環境適応処理のフロー

ソフトウェアの環境適応を実現するため，図 1 の処理フローを提案している．環境適応ソフトウェアは，環境適応機能を中心に，検証環境，商用環境，テストケース DB，コードパターン DB，設備リソース DB の機能群が連携することで動作する．

Step1 コード分析：

Step2 オフロード可能部抽出：

Step3 適切なオフロード部探索：

Step4 リソース量調整：

Step5 配置場所調整：

Step6 実行ファイル配置と動作検証：

Step7 運用中再構成：

ここで，Step 1-7 で，環境適応するために必要となる，コード

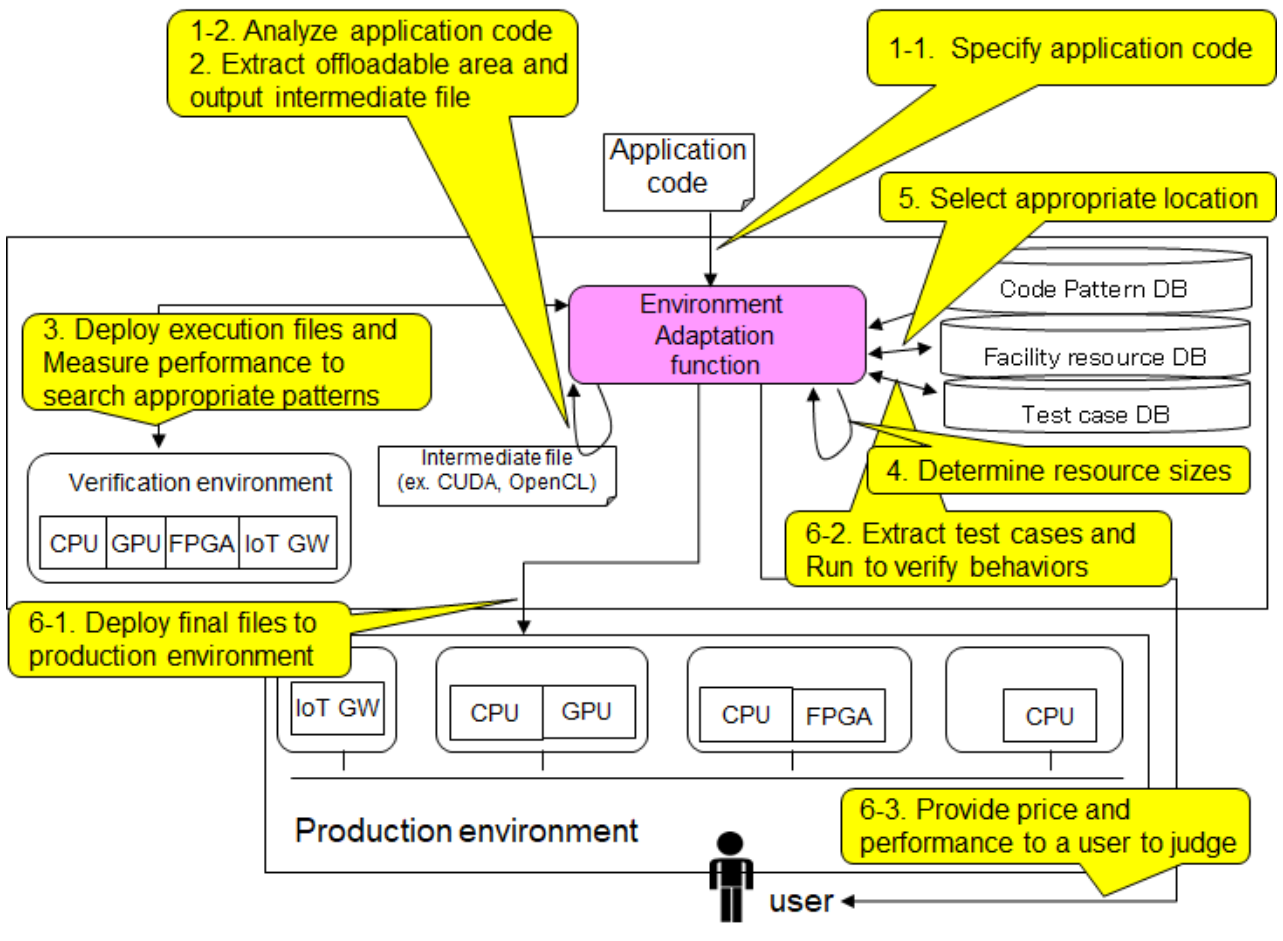

図 1　環境適応ソフトウェアのフロー

の変換，リソース量の調整，配置場所の決定，Jenkins 等 [41] [42] 使った検証，運用中の再構成を行うことができるが，実施したい処理だけ切り出すこともできる．例えば，本稿で対象とする GPU 向けのコード変換だけ実施する場合は，Step 1-3 だけ処理すればよく，利用する機能も環境適応機能や検証環境等だけ利用する形でよい．

### 3.2　以前研究の GPU 自動オフロード手法とその課題

改善元となる，著者の以前の GPU 自動オフロード手法を説明する．大きく 2 つの特徴があり，GPU に適したループ文の抽出を遺伝的アルゴリズム（GA）[43] を用いて行う点を [32] で，ネストループ文で使われる変数をできるだけ上位のループで CPU-GPU 転送を行う点を [33] で提案している．

オフロードしたいアプリケーションは，多様であるが，映像処理のための画像分析（[44] [45] 等），センサデータを分析するための機械学習処理等，計算量，時間が多いアプリケーションでは，ループ文による繰り返し処理が長時間を占めている．そこで，ループ文を GPU に自動でオフロードする事での高速化をターゲットとしている．

まず，基本的な課題として，コンパイラがこのループ文は GPU で並列処理できないという制限を見つけることは可能だが，このループ文は GPU の並列処理に適しているという適合性を見つけることは難しいのが現状である．一般的にループ回数が多い等の計算密度が高いループの方が適していると言われるが，実際に GPU に出すことでどの程度の性能になるかは，実測してみないと予測は困難である．そのため，このループを GPU にオフロードするという指示を手動で行い，性能測定を試行錯誤することが行われている．

[32] はそれを踏まえ，GPU にオフロードする適切なループ文の発見を，GA で自動的に行うことを提案している．並列化を想定していない汎用プログラムから，最初に並列可能ループ文のチェックを行い，次に並列可能ループ文群に対して，GPU 実行の際を 1，CPU 実行の際を 0 と値を置いて遺伝子化し，検証環境で性能検証試行を反復し適切な領域を探索している．並列可能ループ文に絞った上で，遺伝子の部分の形で，高速化可能な並列処理パターンを保持し組み換えていくことで，取り得る膨大な並列処理パターンから，効率的に高速化可能なパターンを探索している．

[33] では，ループ文の適切な抽出に加えて，CPU-GPU 間の転送を減らすため，ネストループ文の中で利用される変数について，ループ文を GPU にオフロードする際に，ネストの下位で CPU-GPU 転送が行われると下位のループの度に転送が行われ効率的でないため，上位で CPU-GPU 転送が行われても問題ない変数については，上位でまとめて転送を行うことを提案している．処理時間がかかるループ回数の多いループは，ネストであることが多いため，転送回数削減による高速化に一定の効果を示す手法である．

この 2 つのアイデアを元に，ループ文が 100 を超える中規模なアプリケーションでも自動高速化を確認しているが，実用性を意識した際に 2 点課題があった．1. より高速化，2. 適用範囲の拡大，である．

より高速化については，OpenACC を用いた自動高速化は，CUDA を用いた手動高速化に比べて，性能改善が十分でないアプリケーションが多いことが言える．CUDA での高速化手法では，逐次処理の並列化は大前提に，CPU-GPU データ転送削減，複数メモリ適切な使い分け（共有メモリ，コンスタントメモリ，テクスチャメモリ，ローカルメモリ，グローバルメモリ），コアレスアクセス，Warp 内分岐の抑制，Warp 同時マルチスレッドによる高 occupancy 化，ストリームによるタスク並列化，スレッド数に適した並列化粒度チューニング等がされる．この中で，転送速度絶対値から効果が大きい点は，GPU 内でのメモリ効率化よりも，CPU-GPU 転送の削減であるため，ネストループ変数以外でも削減できる点を次サブ節で検討する．

適用範囲については，[32] [33] の実装ツールで，進化的計算手法の前に，GPU 向けのコンパイルをする段階でエラーが多発し，高速化が試行できないアプリケーションが幾つかあったことが課題と言える．GPU 処理の指示を行う際に，以前研究ではエラーが出て対象外としていたループ文にも指示を与えられるよう，指示句の拡大を次サブ節で検討する．

### 3.3　CPU-GPU 転送の削減と GPU 処理指定指示句の拡大の検討

CPU-GPU 転送の削減のため，ネストループの変数をできるだけ上位で転送するに加え，多数の変数転送タイミングの一括化，コンパイラが自動転送してしまう転送の削減を検討する．GPU に処理をオフロードするため，CPU-GPU 転送は必ず発生するが，一括化や不要な転送を削減することで，転送数を減らすことでの高速化を実現する．

転送の削減にあたり，ネスト単位だけでなく，GPU に転送するタイミングがまとめられる変数については一括化して転送する．例えば，GPU の処理結果を CPU で加工して GPU で再度処理させるなどの変数でなければ，複数のループ文で使われる CPU で定義された変数を，GPU 処理が始まる前に一括して GPU に送り，全 GPU 処理が終わってから CPU に戻すなどの対応も可能である．コード分析時にループ及び変数の参照関係を把握するため，その結果から複数ファイルで定義された変数について，GPU 処理と CPU 処理が入れ子にならず，CPU 処

理と GPU 処理が分けられる変数については，一括化して転送する指定を OpenACC の data copy 文を用いて指定する．合わせて，一括化して転送され，そのタイミングで転送が不要な変数は data present を用いて明示する．なお，present 節は，既に GPU に変数があることを明示する節である．

コンパイラが自動転送する場合がある転送の削減を検討する．例えば，図 2 は，OpenACC のコンパイラとして著名な PGI コンパイラのケースである．OpenACC の data copy や present 節を用いずに，単にループを#pragma acc kenerls 節で GPU 処理を指定している場合は，ループ単位で CPU と GPU の間でループ内変数の同期が行われる．[33] では，ネストでのこういった問題を減らすため，明示的に data copy を指定していた．しかし，data copy や present を OpenACC で指定した場合でも，コンパイラで変数が CPU-GPU で自動転送される場合がある．コンパイラは基本的に安全側に処理を倒すため，グローバル変数かローカル変数であるか，初期化はどこでされるか，ループ含む他関数から取得されるものか，参照されるだけか，ループ内で更新されるものか，等複数の条件によって，コンパイラ依存で転送が不要であっても転送が発生する．そこで，こういった OpenACC の指示では意図しないが性能を劣化する転送を削減するため，一時領域を作成し一時領域でパラメータを初期化して，CPU-GPU 転送に用いることで，不要な CPU-GPU 転送を遮断する．

次に，適用できるアプリケーション増加のため，指示句の拡大を検討する．具体的には GPU 処理を指定する指示句として，以前研究で用いていた kernels 指示句に加えて，parallel loop 指示句，parallel loop vector 指示句にも拡大する．OpenACC 標準にて，kernels は single loop や tightly nested loop に，parallel loop は non-tightly nested loop も含めたループに，parallel loop vector は parallelize はできないが vectorize はできるループに使われる．tightly nested loop とはネストループにて，例えば，i と j をインクリメントする二つのループが入れ子になっている時，下位のループで i と j を使った処理がされ，上位ではされないような単純なループである．また，PGI コンパイラ等の実装では kernels は並列化の判断はコンパイラが行い，parallel は並列化の判断はプログラマが行うという違いがある．以前の研究では単純なループを対象に検討していたが，non-tightly nested loop や parallelize できないループ等 kernels ではエラーになるループ文は対象外であったため，適用範囲が狭かった．

そこで，本稿では，single, tightly nested loop には kernels を使い，non-tightly nested loop には parallel loop を使い，parallelize できないが vectorize できるループには parallel loop vector を使う．また，parallel 指示句にすることで，結果が kernels の場合より信頼度が下がる懸念があるが，最終的なオフロードプログラムに対して，サンプルテストを行い，CPU との結果差分をチェックしその結果をユーザに見せることで，確認してもらうことを想定している．そもそも，CPU と GPU ではハードが異なるため，有効数字桁数や丸め誤差の違い等があり，kernels だけでも CPU との結果差分のチェックは必要と考えている．

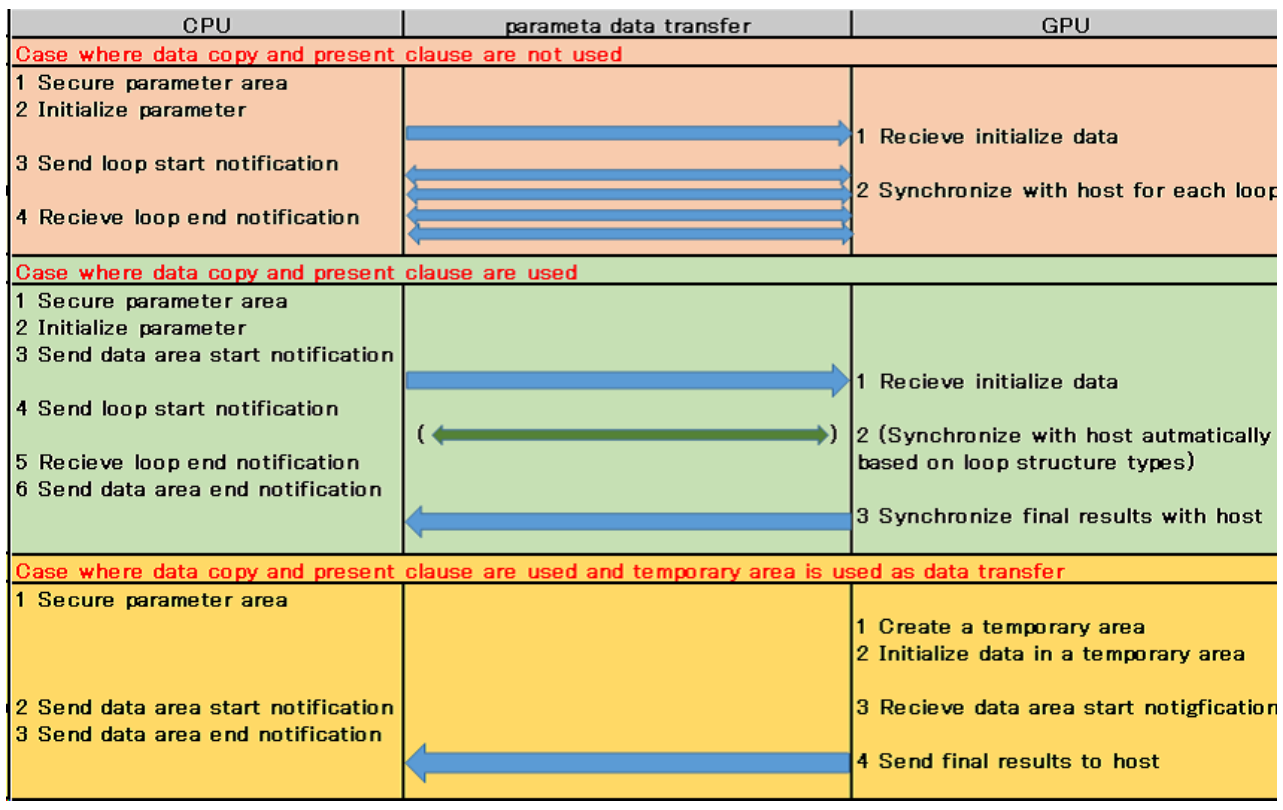

図 2　一時領域を用いた CPU-GPU 転送の削減

## 4. 実　　装

3 節を踏まえて，本稿で提案する方式の実装は，コード分析後，ループ文に対して kernels，parallel loop，parallel loop vector で GPU 処理を指定した個体を準備し，各個体に対して CPU-GPU 転送を減らす一時領域指定・複数ファイルの変数一括転送指定を行い，各個体をコンパイルして検証環境で性能測定を行い，遺伝的アルゴリズムを通じて高速な個体を交叉，突然変異等を行って，より高速なオフロードパターンを試行探索し，最後に定まった解に対して，CPU のみの場合と，GPU を使った場合の計算誤差をサンプルテストを通じてチェックする．

実装は，GPU 自動オフロード改善の有効性確認を目的とし，対象アプリケーションは C/C++言語のアプリケーションとし，GPU 処理は市中の PGI コンパイラ 19.4 を用いる．

PGI コンパイラは OpenACC を解釈する C/C++/Fortran 向けコンパイラであり，for 文等のループ文を，OpenACC のディレクティブ #pragma acc kernels, #pragma acc parallel loop, #pragma acc parallel loop vector で指定することにより，GPU 向けバイトコードを生成し，実行により GPU オフロードを可能としている．合わせて，#pragma acc data copyin/copyout/copy，や#pragma acc data present 等のディレクティブにより，明示的なデータ転送やデータ転送不要の指示が可能である．

実装は C 言語で多くの処理を行い，GA は Perl 5，構文解析関連は Python 2.7 で合わせて処理を行う．

実装は，C/C++アプリケーションの利用依頼があると，まず，C/C++アプリケーションのコードを解析して，for 文を発見するとともに，for 文内で使われる変数データ，その変数の処理等の，プログラム構造を把握する．構文解析には，LLVM/Clang の構文解析ライブラリ (libClang の python binding) [46] を使う．また，CCFinderX [47] 等の類似性検知ツールで含まれている機能ブロック等も把握して良い．

CPU 向けアプリケーションにて，GPU 処理自体が不可な for 文は排除する必要がある．PGI コンパイラの pgcc は，各 for 文に対して，tightly nested loop で kernels 処理可能，paralleize はできないが vectorize はできる等の判定ができる．そこで，そ

のような判定が出た各 for 文に対して，#pragma acc kernels，#pragma acc parallel loop，#pragma acc parallel loop vector ディレクティブ挿入を試行し，コンパイル時にエラーが出るかの判定を行う．もしエラーが出るループ文は，GA の対象外とする．更に，gcov や gprof 等を用いて，ループ回数を把握し，回数が少ないループは対象外にしても良い．

ここで，GPU 処理してもエラーが出ないループ文の数が a の場合，a が遺伝子長となる．遺伝子の 1 は並列処理ディレクティブ有，0 は無に対応させ，長さ a の遺伝子に，アプリケーションコードをマッピングする．

次に，初期値として，指定個体数の遺伝子配列を準備する．遺伝子の各値は，0 と 1 をランダムに割当てて作成する．準備された遺伝子配列に応じて，遺伝子の値が 1 の場合は GPU 処理を指定するディレクティブ #pragma acc kernels，#pragma acc parallel loop，#pragma acc parallel loop vector を C/C++ コードに挿入する．single loop 等は parallel にしない理由としては，同じ処理であれば kernels の方が，PGI コンパイラとしては性能が良いためである．この段階で，ある遺伝子に該当するコードの中で，GPU で処理させる部分が決まる．

次に，Clang で解析した for 文内の変数データの参照関係を元に，data copy, data present, 一時領域の指定を行う．まず，転送が必要なケースは，CPU プログラム側で設定，定義した変数と GPU プログラム側で参照する変数が重なる場合は，CPU から GPU への変数転送が必要であり，GPU プログラム側で設定した変数と CPU プログラム側で参照，設定，定義する変数が重なる場合は，GPU から CPU への変数転送が必要である．転送が必要な中で，GPU で処理した結果を CPU で処理しそれを GPU で再度処理する等の繰り返しがない場合は一括化できる場合がある．具体的には，転送が必要な変数について，GPU 処理の開始前と終了後に一括転送すればよい変数については，複数ファイルで定義された変数を一括#pragma acc data copy で CPU-GPU データ転送を指定する．複数ファイルで一括化できずとも，ネスト単位で一括化できる変数等は従来の [33] 同様一括化する．CPU-GPU 転送を一括化するため，個々のループ文を GPU 処理する際に，既に変数が GPU にある際は，#pragma acc data present を指定して変数転送が不要であることを指定する．また，コンパイラに依存して自動転送で CPU-GPU 転送が生じることを妨げるため，#pragma acc data copy 等で CPU-GPU 転送が必要な際は，GPU 側で一時領域を作成して（#pragma acc declare create) データを初期化し，一時領域を介してデータ同期を行う指示（#pragma acc update）をする．

GPU 計算処理及び CPU-GPU 転送の指示を挿入された C/C++コードを，GPU を備えたマシン上の PGI コンパイラでコンパイルを行う．コンパイルした実行ファイルをデプロイし，サンプルテストツールで性能を測定する．

全個体に対して，テストツール性能測定後，テストツール処理時間に応じて，各個体の適合度を設定する．設定された適合度に応じて，残す個体の選択を行う．選択された個体に対して，交叉処理，突然変異処理，そのままコピー処理の GA 処理を行い，次世代の個体群を作成する．

次世代の個体に対して，指示挿入，コンパイル，性能測定，適合度設定，選択，交叉，突然変異処理を行う．指定世代数の GA 処理終了後，最高性能の遺伝子配列に該当する，ディレクティブ付き C/C++コードを解とする．

最終解に対して，サンプルテストを行い，CPU のみの場合と，GPU を使った場合の計算誤差を，PGI コンパイラの PCAST 機能を用いてチェックする．pgi_compare または acc_compare API のオプション指定で，例えば，IEEE 754 仕様に沿った誤差チェック等各種チェックが可能であり，誤差をユーザに提示し，問題ないか確認頂くことができる．

## 5. 評　　価

### 5.1 評価条件

#### 5.1.1 評価対象

評価対象は，以前研究でも評価した，IoT で多くのユーザが利用すると想定されるフーリエ変換の NAS.FT，及び流体計算の姫野ベンチマークとする．

フーリエ変換処理は，振動周波数の分析等，IoT でのモニタリングの様々な場面で利用されている．NAS.FT [48] は，FFT 処理のオープンソースアプリケーションの一つである．IoT で，デバイスからデータをネットワーク転送するアプリケーションを考えた際に，ネットワークコストを下げるため，デバイス側で FFT 処理等の一次分析をして送ることは想定される．

姫野ベンチマーク [49] は，非圧縮流体解析の性能測定に用いられるベンチマークソフトで，ポアッソン方程式解法をヤコビ反復法で解いている．CUDA 含めて GPU での手動高速化に頻繁に利用されており，自動でも高速化できることの確認のため利用する．

#### 5.1.2 評価手法

GA での高速化の収束については，以前研究 [32] [33] で評価しているため，今回は，GA の各世代に対する，各アプリの性能をグラフでは表示しないが，性能測定では，全個体に対してサンプルアプリでの性能を測定している．指定世代数の試行が終わった際の最高性能コードパターンが，オフロード探索の解であり，その解の性能を，全て CPU 処理に比べて，改善度を評価する．

実行する GA の，パラメータ，条件は以下で行う．

遺伝子長 : 並列可能ループ文数（姫野ベンチマークは 13，NAS.FT は 65．なお，NAS.FT は for 文自体は 82 あるが，GPU 処理できない for 文も多い）

個体数 M : 遺伝子長以下とする（姫野ベンチマークは 10，NAS.FT は 30）

世代数 T : 遺伝子長以下とする（姫野ベンチマークは 10，NAS.FT は 20）

適合度 : $(処理時間)^{-1/2}$　処理時間が短い程高適合度になる．また，(-1/2) 乗とすることで，処理時間が短い特定の個体の適合度が高くなり過ぎて，探索範囲が狭くなるのを防ぐ．また，性能測定が一定時間（3 分）で終わらない際はタイムアウトさせ，処理時間 1000 秒として適合度計算する．

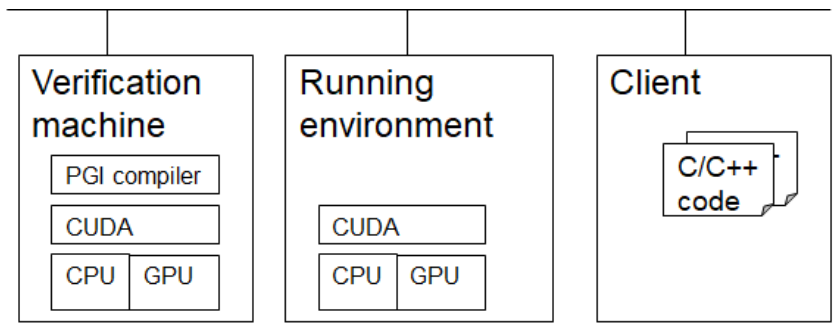

| Name | Hardware | CPU | RAM | GPU | OS | CUDA toolkit | PGI Compiler |
|---|---|---|---|---|---|---|---|
| Verification machine | Dell Precision Tower 3620 | Intel(R) Core(TM) i5-7500 CPU@3.40GHz | 32GB | NVIDIA Quadro P4000 (CUDA core: 1792, Memory: GDDR5 8GB) | Ubuntu 16.04.6 LTS | 10.1 | 19.4 |
| Running environment | Dell Precision Tower 3620 | Intel(R) Core(TM) i5-7500 CPU@3.40GHz | 32GB | NVIDIA Quadro P4000 (CUDA core: 1792, Memory: GDDR5 8GB) | Ubuntu 16.04.6 LTS | 10.1 | |
| Client | HP ProBook 470 G3 | Intel Core i5-6200U @2.3GHz | 8GB | | Windows 7 Professional | | |

図 3　性能測定環境

選択：ルーレット選択．ただし，世代での最高適合度遺伝子は交叉も突然変異もせず次世代に保存するエリート保存も合わせて行う．

交叉率 Pc：0.9

突然変異率 Pm：0.05

### 5.1.3　評 価 環 境

利用する GPU として NVIDIA Quadro P4000 を備えた物理マシンを検証に用いる．NVIDIA Quadro P4000 の CUDA コア数は 1792 である．PGI コンパイラはコミュニティ版の 19.4，CUDA Toolkit は 10.1 を用いる．評価環境とスペックを図 3 に示す．

### 5.2　性 能 結 果

GPU での手動高速化でよく利用されているアプリケーションとして姫野ベンチマーク，多くのユーザが IoT で使うアプリケーションとして，NAS.FT の高速化を確認した．

図 4 は以前研究で高速化を行った際の例である [33]．NAS.FT の，各世代個体の最高性能と GA の世代数をグラフにとり，性能は CPU のみで処理の場合との比で示している．図 4 では，全て 0（全 CPU 処理）の遺伝子では 31.3 秒だったのが，最終的に 5.8 秒で処理し 5 倍の性能が実現出来ていることが分かる．また，GA の中で適応度が高い同じ遺伝子パターンが生じるケースが多いこともあり，7 時間以内でオフロード抽出処理はできた．

以前の結果を踏まえ，今回提案手法の実装により，どの程度性能が改善されたかの測定結果を図 5 に示す．図 5 は，最終解の性能が，全 CPU 処理に比べて何倍になっているかを示した表である．図 5 より，姫野ベンチマークについては，以前研究では 4.8 倍に留まっていたのが，今回提案手法により 15.4 倍の性能が実現できていることがわかる．NAS.FT については，以前研究では 5.4 倍に留まっていたのが，今回提案手法により 10.0 倍の性能が実現できていることがわかる．

## 6.　ま　と　め

本稿では，私が提案している，ソフトウェアを配置先環境に合わせて自動適応させ GPU 等を適切に利用して，アプリケーションを高性能に運用するための環境適応ソフトウェアの要素として，ソフトウェアループ文の GPU へのオフロード自動化手法の改善を提案した．

著者が以前検討した GPU 自動オフロード手法は，進化的計算手法で GPU にオフロードする適切なループ文を抽出し，ネストループ文内の変数をできるだけ上位のループで転送することで，自動高速化を行っていた．以前手法のより高速化，より適用アプリケーション拡大のため，本稿で改善検討を行った．まず，より高速化のため，CPU-GPU 転送を削減することを目的に，一括的に CPU-GPU 転送を行う範囲を複数ファイルの変数にも拡大し，一括転送して GPU に既にある変数は OpenACC の present を用いて転送不要を指示し，かつ変数転送時は一時領域を介して転送する．更に，single loop や tightly nested loop の GPU 処理には OpenACC の kernels を用いて指示し，non-tightly nested loop でも GPU 処理可能なループについては OpenACC の parallel を用いて指示することで，適用できるループ文を拡大した．

以前評価したアプリケーション含め，複数のアプリケーションに対して提案手法での GPU 自動オフロードを行い，方式の有効性を確認した．今後は，より多くのアプリケーションでの評価を行うとともに，ループ文だけでなく，FFT 等大きな機能ブロック単位でのオフロードも検討する．

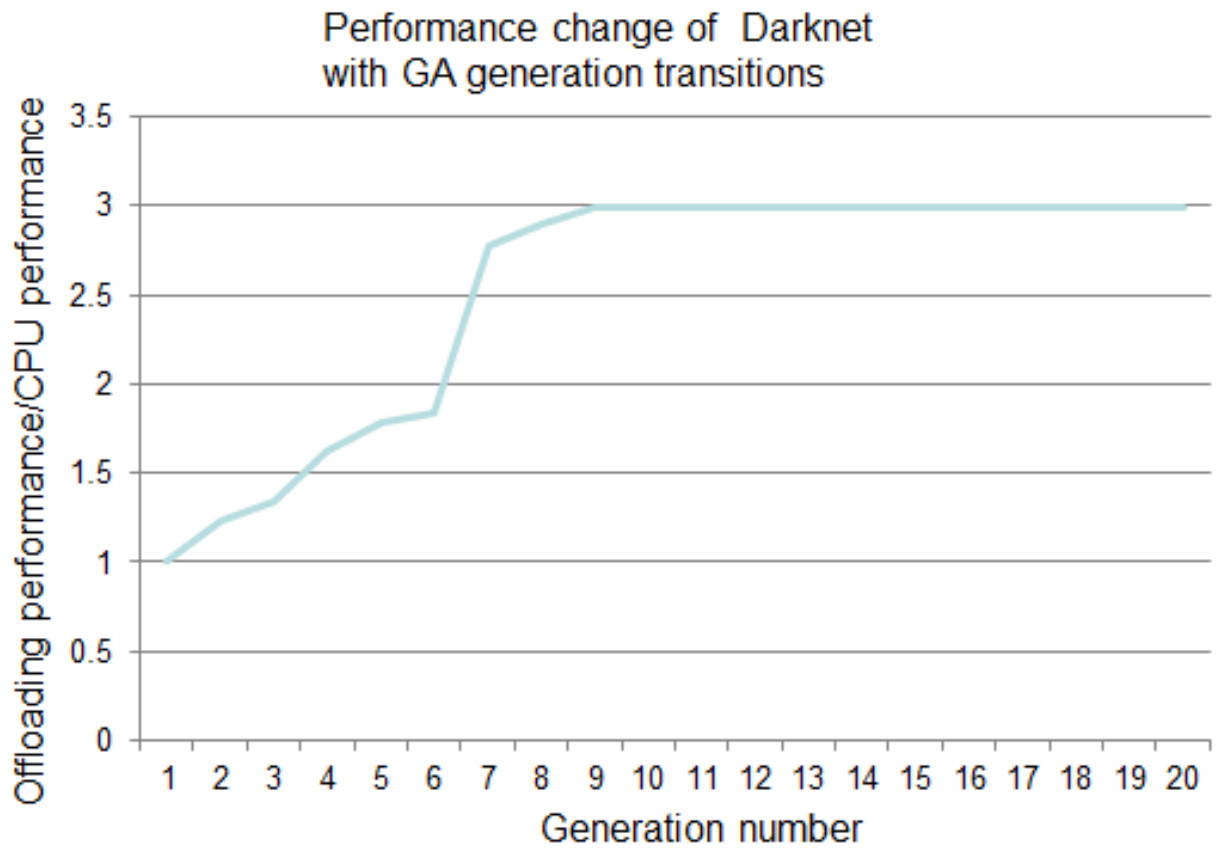

図 4　参考：以前研究での GA 世代数に伴う NAS.FT の性能変化 [33]

| | Performance improvement of previous paper [33] | Performance improvement of this paper proposals |
|---|---|---|
| Himeno benchmark | 4.8 | 15.4 |
| NAS.FT | 5.4 | 10.0 |

図 5　以前研究と今回提案手法での性能改善度の比較

## 文　　献